\documentclass[aps,superscriptaddress,twocolumn,]{revtex4}
\usepackage{bm}
\usepackage{epsfig}
\usepackage{times}
\usepackage{bbm}
\usepackage{amssymb}
\usepackage{amsmath}
\usepackage{natbib}
\usepackage{color}
\usepackage[normalem]{ulem}
\usepackage[latin1]{inputenc}
\usepackage[loose,nice]{units}

\begin{document}
\title{A Schr\"{o}dinger equation for relativistic laser-matter interactions}
\author{Tor Kjellsson Lindblom}
\affiliation{The University of Electro-Communications, 1-5-1 Cholangioma, Choux, Tokyo 182-8585, Japan}
\affiliation{Stockholm University, AlbaNova University Center, SE-106 91 Stockholm, Sweden}
\author{Morten F{\o}rre}
\affiliation{University of Bergen, N-5007 Bergen, Norway}
\author{Eva Lindroth}
\affiliation{Stockholm University, AlbaNova University Center, SE-106 91 Stockholm, Sweden}
\author{S{\o}lve Selst{\o}}
\affiliation{OsloMet -- Oslo Metropolitan University, NO-0130 Oslo, Norway}

\begin{abstract}
A semi-relativistic formulation of
light-matter interaction is derived using the so called propagation gauge and the relativistic mass shift.
We show that relativistic effects induced by a
super-intense laser field can, to a surprisingly large extent, be accounted for by the Schr{\"o}dinger equation, provided that we replace the rest mass in the propagation gauge Hamiltonian by the
corresponding time-dependent
field-dressed
mass.
The validity of the semi-relativistic approach is
tested numerically on a hydrogen atom exposed to an intense XUV laser pulse strong enough to accelerate the electron towards relativistic velocities.
It is found that while the results obtained from the ordinary (non-relativistic) Schr{\"o}dinger equation
generally differ from those of the Dirac equation, demonstrating that relativistic effects are significant,
the semi-relativistic formulation provides results in quantitative agreement with a fully relativistic
treatment.
\end{abstract}

\maketitle

Triggered by rapid technological advances~\cite{Moore1999,DiChiara:08,Yumoto:13,Yoneda:14} and new infrastructure projects~\cite{ELI} there is an increased interest in the description of quantum systems exposed to super-intense laser fields. In spite of the importance to address the relativistic regime~\cite{RevModPhysDiPiazza} there are
comparatively few such
studies reported
in the literature, probably due to the very nature of the time-dependent Dirac equation which is notoriously hard to solve.

Several issues make the Dirac equation tougher to solve numerically than its
non-relativistic counterpart, the Schr{\"o}dinger equation. The fact that the numerical
space is increased by a factor four
owing to the four components of the Dirac wave function, as opposed to a scalar wave
function in the non-relativistic case,
is but the least of problems. The existence of a negative energy continuum is harder to tackle -- for several
reasons~\cite{Selsto2009,Vanne2012}. Firstly, many numerical time integration techniques require a
numerical time step restricted by the inverse of the rest mass energy of the particle at hand, thus
rendering calculations for realistic laser pulses infeasible. Secondly, spurious states may contaminate
the spectrum of the numerical representation of the Hamiltonian.

Another, more subtle complication is the fact that inclusion of the spatial dependence of the external field,
which is imperative for ultrastrong fields~\cite{Reiss2000}, is harder to achieve in a consistent manner for
the Dirac equation than is the case for the Schr{\"o}dinger equation~\cite{kjellsson2017_TDDEvel}. While
this particular challenge to a large extent has been lifted by introducing the so-called {\it propagation gauge}
to the Dirac equation~\cite{simonsen2016,forre:2016, kjellsson2017_PG}, a formulation of the Schr{\"o}dinger
equation which allows us to include relativistic effects is still desirable, albeit seemingly too much to hope for.
However, as it turns out, the apparently
naive
approach of simply substituting the rest mass with the
relativistic mass of the electron, does in fact, to a surprisingly large degree, accommodate for relativistic
effects induced by external electromagnetic fields.

In the following
we outline the theoretical framework. The semi-relativistic interaction is
derived in three ways: first directly from the energy relation
\begin{equation}
\label{RelativisticEnergy}
E= \sqrt{m^2c^4+ p^2 c^2}  ,
\end{equation}
then from the Dirac equation, albeit within the
so-called {\it long wavelength approximation},
and, finally, from the
Klein-Gordon equation.
Our numerical results, comparing the fully relativistic and semi-relativistic approaches, are then presented.
Here we have also included results obtained from
non-relativistic calculations
in order to demonstrate that the studied cases indeed feature relativistic effects. Finally, we present our conclusions.
Atomic units are used where stated explicitly.

First, we will take the following relativistic Hamiltonian for a particle of charge $q=-e$ as our starting point:
\begin{equation}
\label{SqrtHamiltonian}
H_0=
\sqrt{m^2c^4+p^2c^2} - mc^2 - e \varphi(\bf{r})  ,
\end{equation}
which would act on a scalar wave function, as opposed to a bi-spinor in the Dirac case.
For an atom in the absence of any external field, the scalar potential $\varphi$
is simply provided by the Coulomb potential, $V=-e \varphi$.

Of course, both the Dirac and Klein-Gordon formulations are closely related to the above Hamiltonian.
The various ways we present for deriving the
semi-relativistic
form from these formulations are, however,
rather different.
In all cases it is crucial that the interaction is formulated within the propagation
gauge -- for reasons which will be transparent shortly.

We take our external field ${\bf A}$ to be linearly polarized and satisfying the Coulomb gauge restriction,
\begin{equation}\label{Adef}
  {\bf A} = A(\eta) \hat{\bf A} \quad \text{with} \quad \eta = \omega t - {\bf k} \cdot {\bf r} ,
\end{equation}
where the unit vectors $\hat{\bf k}$ and $\hat{\bf A}$ are orthogonal.
The propagation gauge formulation of the interaction is obtained from the usual minimal
coupling formulation by imposing the following gauge transformation~\cite{forre:2016,simonsen2016,kjellsson2017_PG}:
\begin{align}
\nonumber
\bm{A} & \rightarrow \bm{A} +\nabla \xi \quad \text{and} \quad
\varphi \rightarrow \varphi -  \frac{\partial}{\partial t} \xi \quad \text{with} \\
\label{GaugeTrans2}
\xi(\eta) & =
-\frac{e}{2m\omega}   \int_{-\infty}^\eta [A(\eta')]^2 d\eta' .
\end{align}
The corresponding kinetic momentum is now
\begin{equation}
\label{KappaDef}
{\bf d}
= {\bf p} +  e \mathbf{A}+
\frac{e^2}{2mc} A^2 \hat{\bf k} .
\end{equation}
This momentum shift is such that a free, non-quantum mechanical electron starting at zero momentum,
remains at zero momentum. This applies both to the polarization direction and the propagation direction
of the external field, not only to the polarization direction, as is the case in
the usual minimal coupling formulation.

By introducing the interaction with the external field via minimal coupling, ${\bf p} \rightarrow {\bf p} + e {\bf A}$, and
then imposing the gauge transformation Eq.~(\ref{GaugeTrans2}), the Hamiltonian of
Eq.~(\ref{SqrtHamiltonian}) takes the form
\begin{align}
\nonumber
H & =
\sqrt{m^2c^4+\left( {\bf p} + e {\bf A} + \frac{e^2}{2mc} A^2 \hat{\bf k}\right)^2 c^2}
\\ & \nonumber
- mc^2+
V - \frac{e^2}{2m} A^2
 \\ &
\nonumber
=\sqrt{m^2 c^4+d^2 c^2}
- \left(m+\frac{e^2}{2mc^2} A^2 \right) c^2 + V
 \\ &
\label{BeforeExpansion}
=\sqrt{\mu^2 c^4+q^2 c^2}
- \mu c^2 + V  ,
\end{align}
where we have introduced
\begin{align}
\label{MstarDef}
  \mu & = m \left( 1 + \frac{e^2}{2 m^2 c^2} A^2\right) \quad \text{and} \\
\label{Qdef}
  q^2 & = p^2 + 2 e {\bf A}\cdot {\bf p} +
  \frac{e^2}{2 m c} \left\{ A^2,\hat{\bf k} \cdot {\bf p}
\right\}  .
\end{align}
The $\eta$-dependent effective mass $\mu$, which is not to be confused with the reduced
mass, coincides with the
time-dependent relativistic mass of a free, classical electron in the field initially at rest~\cite{Sarachik1970}.
We now expand the square root in a manner which ensures Hermicity term by term:
\begin{align}
\nonumber
H & = V + \frac{c^2}{2} \left( \mu \, \sqrt{1+\mu^{-2}\frac{q^2}{c^2}} + \sqrt{1+\frac{q^2}{c^2} \mu^{-2}} \,  \mu \right) - \mu c^2
\\ &
= V + \frac{1}{2} \left[ \left\{ \frac{1}{2 \mu}, q^2 \right\} - \left\{ \frac{1}{8\mu c^2}, q^2 \mu^{-2} q^2 \right\} + ...\right] .
\label{Expansion}
\end{align}
Truncation at lowest order yields
\begin{equation}
\label{SchrodHamWithMu}
H \approx \frac{1}{2} \left\{ \frac{1}{2\mu}, p^2 \right\} + \frac{e}{\mu} {\bf A} \cdot {\bf p} +
\frac{e^2}{8mc} \left\{ \frac{1}{\mu}, \left\{A^2, \hat{\bf k} \cdot {\bf p} \right\}\right\} + V ,
\end{equation}
where the anti-commutators persist due to the fact that the relativistic mass $\mu$ is spatially dependent and,
hence, does not commute with the momentum operator.

We argue that the leading order term in Eq.~(\ref{Expansion}) indeed includes most of the
relativistic correction to the ionization probability.
Now, one may rightfully question how an expansion of the kinetic energy can meaningfully be truncated at
lowest order in $p^2$ in the relativistic region. This is precisely why it is crucial that the interaction is formulated
in the propagation gauge.
Note that for an initial wave packet with
$\langle {\bf p} \rangle = \boldsymbol{0}$ in the absence of any Coulomb potential,
the expectation value of the canonical momentum will remain identical to zero at all times in the propagation
gauge.
In the presence of the Coulomb potential this is, of course, no longer true, and one can only assume that
$\langle {\bf p} \rangle \simeq 0$. However, in the strong field-limit, the Coulomb potential
eventually represents only a small perturbation, with the result that one can safely neglect
higher order terms in the momentum operator with respect to the leading order term. Thus, we expect that the
validity of the Hamiltonian of Eq.~(\ref{SchrodHamWithMu}) in fact will {\it increase} with increasing
intensity.

In the present approach, relativistic corrections to the Coulomb interaction and spin are not accounted for. However,
we do expect it to correctly accommodate for transient relativistic effects induced by the external laser field as
long as the probability of real pair production is negligible.

The expansion of Eq.~(\ref{Expansion}) becomes simpler
within the long wavelength approximation (LWA), in which the spatial dependence of the vector
potential ${\bf A}$ and, thus, also in $\mu$ is neglected:
\begin{equation}
\label{ExpansionLWA}
H= V + \frac{q^2}{2 \mu} - \frac{q^4}{8\mu^3} + ... \quad .
\end{equation}
If we, again, retain only the leading term in kinetic energy, we obtain
\begin{equation}
\label{SchrHamWithMuLWA}
H \approx \frac{q^2}{2 \mu} + V = \frac{p^2}{2\mu} + \frac{e}{\mu} {\bf A} \cdot {\bf p} + \frac{e^2}{2\mu m c} A^2 \hat{\bf k} \cdot {\bf p} + V .
\end{equation}
We emphasize that the LWA
is far less restrictive than the much applied {\it dipole approximation}.
While the assumption of a homogeneous vector potential imposed within the propagation gauge does
not distort the leading magnetic interaction, no magnetic interaction at all is included with a
homogeneous vector potential imposed within the minimal coupling, or {\it velocity gauge}, formulation.

The Hamiltonians Eqs.~(\ref{SchrodHamWithMu}) and (\ref{SchrHamWithMuLWA}) coincide with the non-relativistic
Schr{\"o}dinger Hamiltonian in the propagation gauge upon the substitution
\begin{equation}
\mu  \rightarrow m .
\end{equation}
Note, however, that the opposite does not hold due to the remaining dependence on the electron rest mass in the third term in Eqs.~(\ref{SchrodHamWithMu}) and (\ref{SchrHamWithMuLWA}). This term is accounting for the leading magnetic interaction in the strong field limit and its rest-mass dependence originates from the momentum induced in the direction of the electromagnetic field, i.e., the third term on the right-hand side of Eq.~(\ref{KappaDef}).

It is worth mentioning that the interaction form of Eq.~(\ref{SchrHamWithMuLWA}) may rather easily be transformed into other more familiar forms, analogous to, e.g., the velocity gauge, the length gauge or the Kramers-Henneberger frame \cite{Selsto2007,Bandrauk2013}.

Next, we will outline how the above Hamiltonian may be derived from the Dirac equation.
In propagation gauge-form it reads~\cite{kjellsson2017_PG}
\begin{align}
\nonumber
& i \hbar \frac{d}{dt} \Psi = H \Psi \quad \text{with} \\
\label{PGhamiltonian}
&H  =
c \boldsymbol{\alpha} \cdot
{\bf d}
  + m c^2 \beta  +
  \left( V - \frac{e^2}{2 m}A^2 \right) \mathbbm{1}_4 ,
\end{align}
where
${\bf d}$
is
defined in Eq.~(\ref{KappaDef}).
The wave function $\Psi$ now is a four-component bi-spinor,
\begin{equation}
\Psi=\left( \begin{array}{c} \Phi \\ X  \end{array}\right) .
\end{equation}
The upper spinor $\Phi$ is typically referred to as the {\it large}
component for states with positive energy, while the lower spinor, $X$, is coined the {\it small} component.
We apply the usual formulation in terms of Pauli-matrices, $\boldsymbol{\sigma}$, and
identity matrices for the $\boldsymbol{\alpha}$ and $\beta$ matrices.

In a classic paper~\cite{foldy:50} Foldy and Wouthuysen showed how the Dirac equation for a free fermion
may be transformed in a manner which decouples the large and the small component.
Specifically, by applying the unitary transformation
\begin{equation}
\label{FWtrans}
T=e^S \quad \text{with} \quad S= \frac{1}{2p}
\tan^{-1} \left( \frac{p}{mc}  \right) \beta \boldsymbol{\alpha} \cdot {\bf p} ,
\end{equation}
the field free Hamiltonian $c \boldsymbol{\alpha} \cdot {\bf p}  + mc^2 \beta$ is cast into
\begin{equation}
\label{FWhamFree}
e^S \left(c \boldsymbol{\alpha} \cdot {\bf p} + mc^2 \beta \right) e^{-S} =
\beta \sqrt{m^2 c^4  +  p^2 c^2 } .
\end{equation}
As may be confirmed by inspection, within the
LWA
the above still holds with the substitution
${\bf p} \rightarrow
{\bf d}
$.
However, since
${\bf d}$ in Eq.~(\ref{KappaDef})
unlike ${\bf p}$, contains explicitly time-dependent terms, the Hamiltonian resulting from imposing the
transformation Eq.~(\ref{FWtrans}) on Eq.~(\ref{PGhamiltonian}) acquires an additional term given by
the time derivative of the
kinetic
momentum:
\begin{align}
\label{FWtransformOnH}
H' & = e^S H e^{-S} + i \hbar \frac{d}{dt} S =
\nonumber  \\
&
\beta \sqrt{m^2 c^4 +
d^2
c^2} + V -\frac{e^2}{2m} A^2 +
i \frac{\hbar}{2} \beta \boldsymbol{\alpha} \cdot \boldsymbol{\lambda}
\\
\text{where} \, \,
&
\label{LambdaDef}
\boldsymbol{\lambda} = \frac{d}{dt}
\frac{{\bf d}}{d}
\tan^{-1} \left( \frac{d}{mc} \right) .
\end{align}
Here we have, again, neglected relativistic corrections arising from
the Coulomb potential, $e^S V e^{-S} \approx V$, i.e.,
contributions such as the Darwin term
and spin-orbit coupling are unaccounted for.

The last term in Eq.~(\ref{FWtransformOnH}), which contains the $\boldsymbol{\alpha}$-operators, re-introduces
coupling between $\Phi$ and $X$. It is, however, negligible in most cases. In order to see this, we rewrite the
Dirac equation, Eq.~(\ref{PGhamiltonian}),
with the transformed Hamiltonian Eq.~ (\ref{FWtransformOnH}) in
terms of the two components:
\begin{align}
\label{Decouple}
& i \hbar \frac{d}{dt}
\left( \begin{array}{c} \Phi \\ X\end{array} \right)  = H'
\left( \begin{array}{c} \Phi \\ X\end{array} \right)
\quad \text{with}
\\
\nonumber
& H' = \left( \begin{array}{cc}
\sqrt{\mu^2 c^4 + q^2 c^2} &
i \frac{\hbar}{2} \boldsymbol{\sigma} \cdot
\boldsymbol{\lambda}
\\
-i \frac{\hbar}{2} \boldsymbol{\sigma} \cdot
\boldsymbol{\lambda}
&
-\sqrt{\mu^2 c^4 + q^2 c^2}
\end{array} \right)
\\
\nonumber
& + \left( V-\mu c^2\right) \mathbbm{1}_4 .
\end{align}
Here we have shifted the energy downwards by $mc^2$ and used the same steps as in connection with Eq.~(\ref{BeforeExpansion}).
We now approximate
\begin{equation}
\label{DeCoupAssumption}
\left[ -\sqrt{\mu^2 c^4 + q^2 c^2} -\mu c^2+ V -i \frac{d}{dt}\right] X \approx
-2\mu c^2 X ,
\end{equation}
i.e., we rely on that in the
propagation gauge the effect of the operator $q^2$
is
small.
With this, Eq.~(\ref{Decouple}) decouples and we are left with the following effective Hamiltonian for the large component $\Phi$:
\begin{equation}
\label{DiracNesteFerdig}
\sqrt{\mu^2 c^4 + q^2 c^2} -\mu c^2+ V + \frac{\hbar^2}{8\mu c^2}
\lambda^2 .
\end{equation}
Here we notice that the three first terms are identical to the right-hand side of Eq.~(\ref{BeforeExpansion}). The remaining term, $\sim \lambda^2$, is to leading order a purely time-dependent one, which can be removed by a phase transformation.
The next to leading term is of order $m^{-4} c^{-5}$ and thus the last term of Eq.~(\ref{DiracNesteFerdig}) may safely be
neglected in view of the approximations already made when using Eq.~(\ref{DeCoupAssumption}).
With this the Hamiltonian of Eq.~(\ref{BeforeExpansion}) is reproduced for the large component $\Phi$, albeit within the
LWA.
We emphasize that the equivalence with the scalar Hamiltonian of Eq.~(\ref{BeforeExpansion})
implies the neglect of all spin-dependent interactions.

Finally, we will outline how Eq.~(\ref{BeforeExpansion}) may be obtained from the Klein-Gordon equation.
Within the propagation gauge, it takes on the
form
\begin{align}
\label{KGequation}
&\left[ -\hbar^2\frac{d^2}{dt^2} + i \hbar \frac{d}{dt} K + K i \hbar \frac{d}{dt}\right] \Psi
= L \Psi ,
\end{align}
where we have
defined the operators
\begin{align}
\label{Adef3}
& K = m c^2 + \frac{e^2}{2m} A^2 -V \quad \text{and} \\
\label{Bdef}
& L = p^2 c^2 +
 \left( e{\bf A} + \frac{e^2}{2mc} A^2 \hat{\bf k} \right) \cdot {\bf p} \, c^2+
\\
&
\nonumber
  {\bf p} \cdot \left( e{\bf A} + \frac{e^2}{2mc} A^2 \hat{\bf k} \right)c^2
+ V \left( 2mc^2+ \frac{e^2}{m}A^2 - V\right) ,
\end{align}
and, again, shifted the energy downwards by the rest mass energy $mc^2$.
By insisting that the Hamiltonian $H$ equals $i \hbar \; d/dt$, we arrive at
\begin{equation}
\left(H^2 + HK + KH - L \right) \Psi=0 .
\end{equation}
The operator on the left hand side above is identical to the zero-operator if and only if
\begin{equation}
\label{KGsolution}
H= - K \pm \sqrt{K^2+L}  .
\end{equation}
Note that this Hamiltonian, just like the Dirac Hamiltonian, provides both a positive and a negative energy spectrum.
By selecting the form corresponding to positive energies we, again, arrive at Eq.~(\ref{BeforeExpansion}),
as may be verified by inspection.

Having derived a Schr{\"o}dinger-like equation for
the
relativistic laser-matter
interactions, we
now aim at demonstrating
its capabilities by performing
a calculation on a concrete
time-dependent problem
and compare its result with the exact one as obtained by the Dirac
equation. To this end, will investigate the following scenario:
A hydrogen atom with the electron initially prepared in
the ground state is exposed to a laser pulse, defined as
\begin{equation}
\label{Adef2}
{\bf A}(\eta) =  A(\eta) \hat{\bf z} = \frac{E_0}{\omega} f(\eta) \sin(\eta + \phi) \, \hat{\bf z}  ,
\end{equation}
where the envelope function $f$ is chosen to be a sine squared.

Both the Dirac equation and the Schr{\"o}dinger equation are solved within the LWA. In the latter
case calculations have been performed both with and without the relativistic mass shift included in order to see whether this shift
can accommodate for relativistic effects. For details on the implementation,
see Refs.~\cite{kjellsson2017_TDDEvel,kjellsson2017_PG}.

\begin{figure}
  \centering
   \includegraphics[width=7.5cm]{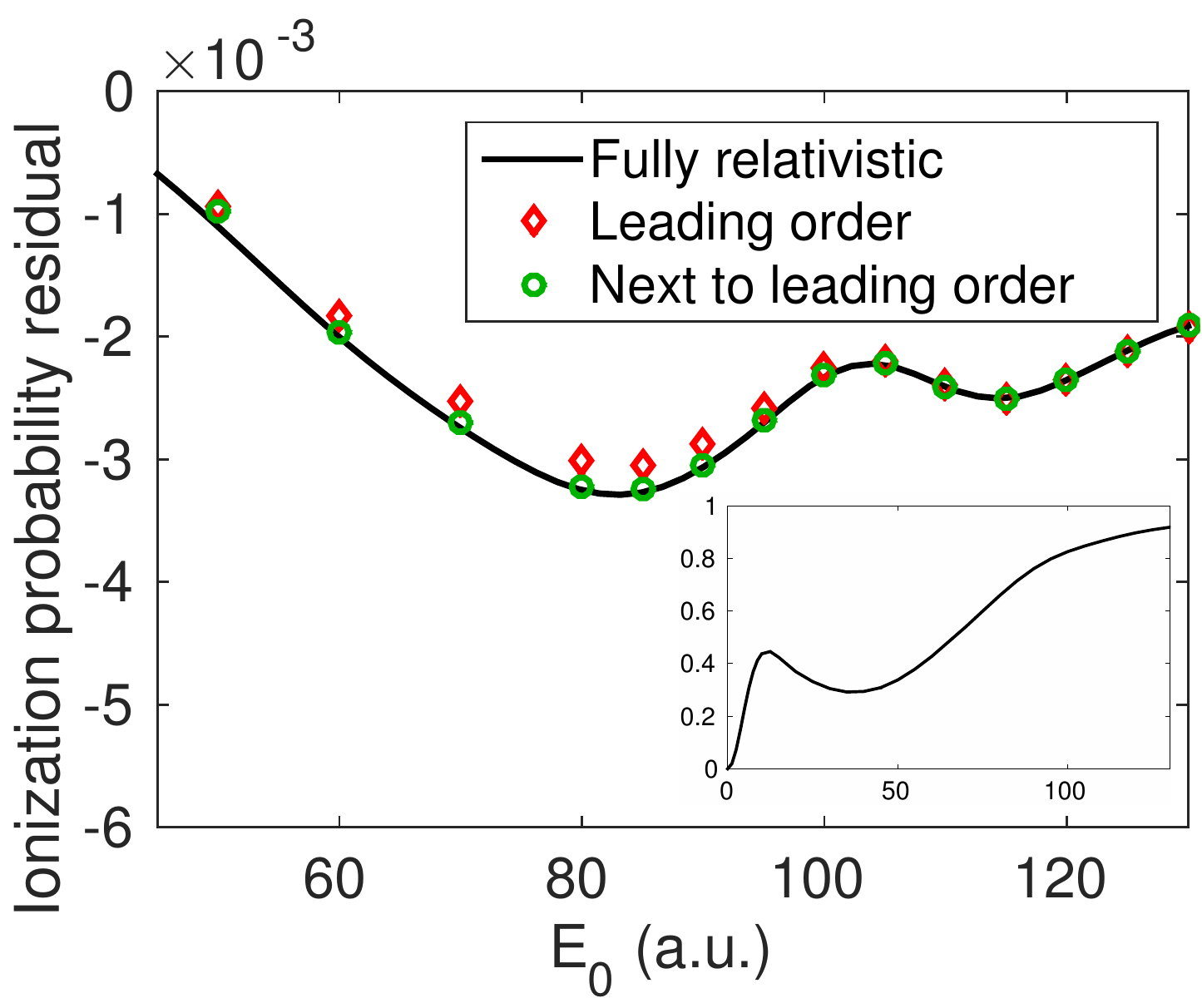}
  \caption{A hydrogen atom initially prepared in the ground state is exposed
  to a 15 cycle laser pulse with a central frequency of
  $\omega=3.5$~a.u. The peak field strength $E_0$ ranges from $50$ to $130$~a.u.
  The black curve is the \textit{difference} in ionization probability predicted by the fully relativistic Dirac equation and the
  non-relativistic Schr{\"o}dinger equation, within the long wavelength approximation.
  The red diamonds show the corresponding difference for results obtained using the semi-relativistic approach with the kinetic energy truncated at lowest order, Eq.~(\ref{SchrHamWithMuLWA}),
  while the green circles are obtained with the next to leading order correction in Eq.~(\ref{ExpansionLWA}) included. The insert shows the {\it total} ionization probability.} 	
  \label{ResFigQM}
\end{figure}

The results for a 15 cycle laser pulse with photon energies in the extreme ultraviolet
region are shown in Fig.~\ref{ResFigQM}. We find relativistic corrections of about 0.5~\% in the ionization
probability. We see that the bulk of the correction is indeed provided by Eq.~(\ref{SchrHamWithMuLWA}), while
the next order corrections from Eq.~(\ref{ExpansionLWA}) in fact vanishes beyond maximum field strengths $E_0$ of about 100~a.u.

\begin{figure}
  \centering
  \includegraphics[width=8cm]{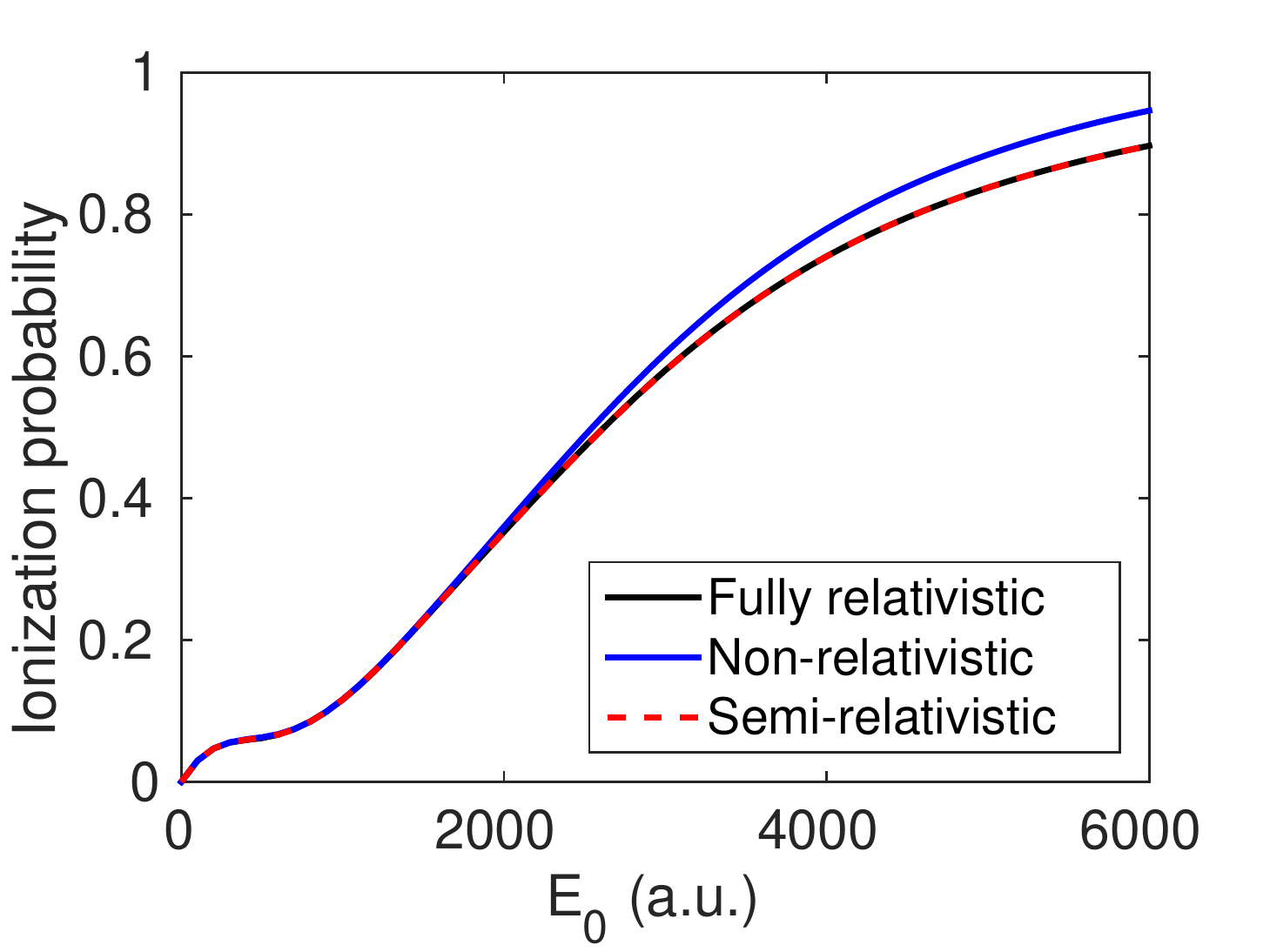}
  \caption{The ionization probability as a function of peak electric field strength, $E_0$,
  obtained by classical trajectory Monte Carlo calculations. The atom has been exposed to a 15 cycle laser pulse with central frequency $\omega=50$~a.u. The calculations have been done both
  fully relativistically, fully non-relativistically and by using the semi-relativistic Hamiltonian function of Eq.~(\ref{SchrHamWithMuLWA})
  with the spatial dependence of the laser field included.}
  \label{ResFigCl}
\end{figure}

Beyond the validity of the
LWA,
we have performed classical trajectory Monte Carlo simulations based on non-quantum mechanical
Hamiltonian functions corresponding to the same three cases as above: fully relativistic,
fully non-relativistic and semi-relativistic. In the semi-relativistic case, our Hamiltonian function takes the same form as in Eq.~(\ref{SchrHamWithMuLWA}), albeit with full spatial dependence in
${\bf A}$ and $\mu$ in this case.
In Fig.~\ref{ResFigCl} we present ionization probabilities obtained from a calculation with a central frequency of
$\omega=50$~a.u., which corresponds to photons in the soft X-ray region.
As we can see, the semi-relativistic results are
virtually indistinguishable from the fully relativistic ones, thus providing strong evidence of the
validity of the present approach.

We have also calculated classical trajectories for electrons, initially at rest, only subject to the interaction with the external laser field, i.e., without any Coulomb interaction. This was done for a wide range of laser frequencies $\omega$, ranging from the infrared region via the optical to the X-ray region.
The same three approaches as above were applied.
While the non-relativistic Hamiltonian produced results deviating substantially from the fully relativistic one, the semi-relativistic formalism consistently gave trajectories virtually indistinguishable from the fully relativistic ones.

In conclusion,
we have demonstrated that by substituting the mass with the {\it relativistic} mass in the adequate manner,
the validity of the Schr{\"o}dinger equation is extended into the relativistic region.
The validity of the approach has been demonstrated by direct comparison with fully relativistic calculations, providing quantitative agreement. This high degree of accuracy is a very surprising finding, and a very useful one indeed:
within the long wavelength approximation, adjustment of the mass is easily implemented, thus extending
the validity of theoretical studies considerably with little extra effort when it comes to implementation and numerics.

As a next step, a derivation based on the Dirac equation beyond the long wavelength approximation will,
hopefully, provide a relativistic Hamiltonian of Schr{\"o}dinger type which also features spin dynamics.

\section*{Acknowledgments}
Simulations have been performed on resources provided by the Swedish National Infrastructure for Computing (SNIC) and on resources provided by the Norwegian Metacenter for Computational Science (UNINETT Sigma2, project number NN9417K). Financial support by the Swedish Research Council (grant number 2016-03789) is also gratefully acknowledged.

\end{document}